\documentclass[]{jkas} % Option for manuscript
\usepackage{siunitx}
\usepackage{mathtools}
\usepackage{threeparttable}
\usepackage{rotating}

\usepackage[utf8]{inputenc}
\usepackage{newtxtext,newtxmath}

\def\year{1919} % publication year
\def\volume{5,381} % journal volume
\def\issue{1,222,320} % journal issue
\def\beginpage{1} % first page of article
\def\received{July 4, 2026} % date paper was received by JKAS (ex. October 12, 2022)
\def\accepted{July 28, 2026} % date of acceptance (ex. December 20, 2022)
\def\published{---} % date of publication (ex. January ??, 2023)
\date{Received \received; Accepted \accepted; Published \published}

\title{%
An Enhanced Catalog of Gaia DR3 Galaxy Candidates with Spectroscopic and Machine-Learning Photometric Redshifts
}

\author[1]{Junghyun Hwang}{0009-0006-8338-2308}
\author[1,2,3]{Ho Seong Hwang}{0000-0003-3428-7612}
\affil[1]{Astronomy Program, Department of Physics and Astronomy, Seoul National University, Gwanak-gu, Seoul 08826, Republic of Korea}
\affil[2]{SNU Astronomy Research Center, Seoul National University, Gwanak-gu, Seoul 08826, Republic of Korea}
\affil[3]{Institute for Data Innovation in Science, Seoul National University, Seoul 08826, Korea}
\def\corrauthor{%
Ho Seong Hwang
\email{hhwang@astro.snu.ac.kr}
}

\def\runningauthor{%
Hwang \& Hwang
}

\def\runningtitle{%
An Enhanced Catalog of GAIA DR3 Galaxy Candidates %%% END %%%%%%%%%%%%%%%%%%%%%%%%%%%%%%%%%%%%%%%%%%%%%%%%%%%%%%%%%%%%%
}

\def\keywords{%
Catalogs (205); Redshift surveys (1378); Neural networks (1933); Astrostatistics (1882); Bayesian statistics (1900)
}

\def\abstracttext{%
We present an updated galaxy catalog based on the Gaia Data Release 3 (DR3) Galaxy Candidates table, augmented with both spectroscopic and machine-learning based photometric redshift estimates. While the original catalog of Gaia DR3 galaxy candidates provides an unprecedented all-sky sample of extragalactic sources, its scientific utility for extragalactic and cosmological studies has been limited by the absence of redshift information and by its relatively low source purity.
To address these limitations, we cross-match the catalog of Gaia DR3 galaxy candidates with existing spectroscopic redshift catalogs including DESI, SDSS, 2MRS, and NED. For sources without spectroscopic measurements, we compute photometric redshifts using a Multiple-Bin Regression Neural Network (MBRNN) machine-learning model. The input features for this model are drawn from a newly constructed multiwavelength catalog that combines Gaia DR3 measurements with 2MASS and unWISE photometry, as well as shape parameters, QSO classification flags, and extinction values. The model is trained and validated using the spectroscopic subsample.
As a result, spectroscopic redshifts are assigned to 24.5\% of the catalog sources, while photometric redshifts are provided for the remaining objects. The MBRNN model achieves improved redshift accuracy and precision compared to previously published photometric redshift estimates based on Gaia data alone. In addition to redshift values, we provide a score quantifying the likelihood that a given source represents a statistical outlier.
The resulting catalog significantly enhances the scientific value of the Gaia DR3 galaxy candidates sample and provides a valuable resource for large-scale extragalactic and cosmological studies.
}

\begin{document}
\jkashead
%%%%%%%%%%%%%%%%%%%%%%%%%%%%%%%%%%%%%%%%%%%%%%%%%%%%%%%%%%%%%%%%%%%%%
%%% BEGIN MAIN TEXT HERE %%%%%%%%%%%%%%%%%%%%%%%%%%%%%%%%%%%%%%%%%%%%
%%%%%%%%%%%%%%%%%%%%%%%%%%%%%%%%%%%%%%%%%%%%%%%%%%%%%%%%%%%%%%%%%%%%%

\section{Introduction}
An accurate measurement of galaxy redshifts is fundamental to modern extragalactic astronomy and cosmology \citep{Peacock2001}.
In this regard, extracting meaningful statistical constraints requires galaxy samples that satisfy several key requirements \citep{Colless2001, Heintz2020, Hwang2016}.
These include the availability of redshift information, which allows distance estimation; wide sky coverage, enabling comprehensive sampling of the universe without introducing directional bias; and a level of homogeneity, which is often achieved by imposing magnitude limits or other well-defined selection functions. 
Large surveys such as the Dark Energy Spectroscopic Instrument (DESI)  \citep{DESI2025}, the Sloan Digital Sky Survey (SDSS) \citep{SDSSDR17}, and the Two
Micron All-Sky Survey (2MASS) Redshift Survey (2MRS) \citep{2MASS2006, 2MRS2012}, and the A-SPEC survey \citep{Kwon2026} have been designed to meet these considerations and thus serve as excellent resources for modern extragalactic astronomy and cosmology.

Although ground-based surveys have provided much of the foundational data for these studies, space-based observatories bring additional capabilities that can significantly enhance the quality of astronomical catalogs. 
\textit{Gaia}, being a space-based telescope, exemplifies this with its unparalleled precision in astrometric and photometric measurements \citep{GaiaCollaboration2016}.
The space environment and dedicated instrumentation of \textit{Gaia} enable a unique combination of high accuracy, high sensitivity, wide dynamic range and full-sky coverage. 
This makes it possible to identify and characterize sources across a broad range of magnitudes and angular positions. 
\textit{Gaia} was primarily designed for the distance and velocity measurements of stars within the Milky Way, but its imaging depth and coverage have also made it a useful tool for detecting extragalactic sources. 
These sources are listed in the \textit{Gaia} DR3 \texttt{galaxy\_candidates} table, which includes 4,842,342 sources classified as the potential extragalactic objects \citep{GaiaCollaboration2023}.
Unlike dedicated extragalactic surveys, a large proportion of the galaxies in this catalog lack redshift measurements, which limits their immediate applicability for cosmological studies.

Furthermore, \citet{GaiaCollaboration2023} showed that the purity of this catalog is only around 50-70\%.
This is because the selection procedure was focused mainly on the completeness of the table rather than its purity. 
This further increases the need for distance measurements, which is often derived from the redshift.

To enhance the scientific utility of the \textit{Gaia} DR3 \texttt{galaxy\_candidates} table, this study aims to incorporate both spectroscopic redshifts (spec-$z$) and photometric redshifts (photo-$z$), thus enabling redshift-dependent analyses, including distance estimation.
Spec-$z$ will be incorporated by cross-matching the \textit{Gaia} extragalactic sample to some external spec-$z$ catalogs, while photo-$z$ will be estimated via machine learning (ML) techniques. 

There are a number of approaches to measuring photo-$z$ using ML techniques, reflecting the diversity and adaptability of ML techniques in astronomy. 
Broadly, these methods fall into supervised learning frameworks, where models are trained on spectroscopic redshifts and corresponding photometric data. 
Studies such as \citet{Firth2003} and \citet{Collister2004} demonstrated that photo-$z$ can be calculated using artificial neural networks with photometry data as input.
Since then, numerous machine learning algorithms, such as random forest, $k$-nearest neighborhood have been implemented in calculating the photo-$z$ (e.g. \citet{Carliles2010, Ball2007, Salvato2019, MBRNN, Kim2025}).
In this work, a neural network-based ML model called MBRNN will be used to estimate the photo-$z$ of \textit{Gaia} DR3 \texttt{galaxy\_candidates} sources.
Moreover, to address the limited photometric information available in \textit{Gaia}, the input features will be supplemented with photometric data from the 2MASS and unWISE catalogs.

The goal of this study is to produce a data set with complete redshifts, which retains the spatial and photometric advantages of \textit{Gaia} while overcoming its limitations in redshift coverage.
Furthermore, an additional step will be included to assign a probabilistic score to each source, indicating the likelihood of being a genuine galaxy. 
This probability metric, based on Bayesian inference and model uncertainty, adds a layer of reliability that is crucial for filtering and statistical modeling. 
Together, these improvements aim to transform the \textit{Gaia}-derived catalog into a more robust tool for a wide range of astrophysical and cosmological applications.

\section{Data}
\subsection{Spectroscopic Redshifts}\label{specz}
We first assign spec-$z$ to the \textit{Gaia} DR3 \texttt{galaxy\_candidates} table. 
These spec-$z$ values are not only valuable in their own right, but are also essential for training, validation, and testing steps of the ML pipeline.
They are obtained from a combination of several external catalogs, including the DESI DR1 \citep{DESI2025}, SDSS DR17 \citep{SDSSDR17},
2MRS \citep{2MRS2012}, and NASA/IPAC Extragalactic Database (NED)\footnote{\url{https://ned.ipac.caltech.edu/}}. 

For the DESI catalog, both the bright- and dark-time target files of the \texttt{zcatalog} satisfying \texttt{ZWARN == 0} are used.
The SDSS DR17 catalog is further supplemented with redshift data from the literature (see \citet{Bahk2024, Hwang2010, Hwang2014} for details). 
The 2MRS catalog is similarly supplemented with redshift data from various literature;
the redshifts of the sources without measurements in the 2MASS External Source Catalog (XSC;  \citealt{2MASSXSC}) are adopted from  \citet{Quintana1995}, UZC \citep{Falco1999}, PSCz \citep{Saunders2000}, 2dFGRS \citep{Colless2001}, 6dFGRS \citep{Jones2004}, and SDSS DR17 \citep{SDSSDR17}. 
The NED catalog is implemented by querying sources in NED \texttt{objdir} whose \texttt{pretype} field is either \texttt{'G'} or \texttt{'*'} and the redshift is obtained by spectroscopic methods (\texttt{zflag='SCO' OR zflag='S1L' OR zflag='SML' OR zflag='SMU' OR zflag='SPA' OR zflag='SSE' OR zflag='SST' OR zflag='SSN' OR zflag='SUN' OR zflag='STO'})
The query was conducted in July 2025.

A source in \textit{Gaia} DR3 \texttt{galaxy\_candidates} is considered matched  to a source in redshift catalogs if the angular distance between the two positions is less than a catalog-specific threshold, $\Delta\theta_{\max}$.
The summary of selection and cross-matching criteria is presented in the upper section of Table~\ref{ExtTab}, as well as the summary of the different external spectroscopic redshift catalogs used.
In cases where a source is matched to more than one redshift catalog, we adopted a single spectroscopic redshift following a predefined priority order, favoring DESI DR1 over SDSS DR17, followed by 2MRS and finally NED.

\begin{table*}[!t]
\centering
\caption{
External catalogs for spectroscopic redshift and photometric data.
}
\label{ExtTab}
\begin{threeparttable}
\begin{tabular}{c l l r c l}
\toprule
No. & Catalog & Reference & $N_{\text{matched}}$\tnote{a} & $\Delta\theta_{\max}$ (") & Selection / Notes \\
\midrule
\multicolumn{6}{l}{\textit{Spectroscopic Redshift Catalogs}} \\
\midrule
1 & DESI DR1
  & \citet{DESI2025}
  & 625{,}447
  & 0.1
  & \texttt{ZWARN == 0} \\

2 & SDSS DR17
  & \citet{SDSSDR17}
  & 579{,}681
  & 0.3
  & Redshifts supplemented from literature\tnote{b}. \\

3 & 2MRS
  & \citet{2MRS2012}
  & 433{,}107
  & 1.5
  & Redshifts supplemented from literature\tnote{c}. \\

4 & NED query
  & ---
  & 715{,}424
  & 1.0
  & Selected \texttt{zflag}\tnote{d} and \texttt{pretype}\tnote{e}. \\

\midrule
\multicolumn{6}{l}{\textit{Photometric Catalogs}} \\
\midrule
1 & 2MASS XSC
  & \citet{2MASSXSC}
  & 1{,}009{,}460
  & 1.5
  & $J, H, K_s$ 20~mag~arcsec$^{-2}$ isophote mag. \\

2 & unWISE
  & \citet{unWISE2019}
  & 4{,}274{,}296
  & 2.0
  & $W1, W2$. \\

\bottomrule
\end{tabular}

\begin{tablenotes}
\item[a] $N_{\text{matched}}$ denotes the number of sources matched to the \textit{Gaia} DR3
\texttt{galaxy\_candidates} table.
\item[b] \citet{Bahk2024, Hwang2010, Hwang2014}
\item[c] \citet{Quintana1995, Falco1999, Saunders2000, Colless2001, Jones2004, 2MRS2012, SDSSDR17, Hwang2010}
\item[d] \texttt{`SCO', `S1L', `SML', `SMU', `SPA', `SSE', `SST', `SSN', `SUN', `STO'}
\item[e] \texttt{`G', `*'}
\end{tablenotes}
\end{threeparttable}
\end{table*}

\subsection{Input Data for Photometric Redshift Estimation}\label{inputdata}
We then construct a table upon which the ML model would train and estimate photo-$z$, by implementing photometry and supplementary data to \textit{Gaia} DR3 \texttt{galaxy\_candidates} table.
\textit{Gaia} provides photometry in three broad bands within the visible wavelength range \citep{GaiaCollaboration2016}: $G_\mathrm{BP}$ (532 nm), $G$ (673 nm), and $G_\mathrm{RP}$ (797 nm) \citep{Jordi2010}.
However, this limited spectral coverage is often insufficient for training ML models to accurately estimate redshifts. 
To supplement this, additional photometric data are incorporated in this study from external catalogs such as the 2MASS XSC \citep{2MASSXSC}, which covers the near-infrared with three passbands \textit{J} (1.25 $\mu$m), \textit{H} (1.65 $\mu$m), and \textit{K}$_s$ (2.16 $\mu$m) \citep{2MASS2006}; and the unWISE catalog of the wide-field infrared survey explorer mission, which extends into the mid-infrared with two passbands \textit{W1} (3.4 $\mu$m) and \textit{W2} (4.6 $\mu$m) \citep{unWISE2019}.
Therefore, eight photometric bands are implemented in total.

These photometric data are compiled by cross-matching the sources in the 2MASS XSC and unWISE catalogs with the sources in \textit{Gaia} DR3 \texttt{galaxy\_candidates} using a custom \texttt{python} script.
As in Section \ref{specz}, the external catalogs are matched with \textit{Gaia} DR3 \texttt{galaxy\_candidates}, if the angular separation is smaller than some catalog-specific $\Delta\theta_{\max}$.
This threshold varies by catalog because the point spread function depends on factors such as aperture sizes and observational wavelengths.
The summary of the external photometry catalogs and the adopted values of $\Delta\theta_{\max}$ are listed in the lower section of Table~\ref{ExtTab}.

All fluxes or magnitudes are converted to AB magnitudes if they are not provided in AB magnitude. For \textit{Gaia} photometry, the conversion formula and zero point value from section 5.4.1 of the \textit{Gaia} DR3 \hyperlink{https://gea.esac.esa.int/archive/documentation/GDR3}{Online Documentation} is applied to \texttt{phot\_g\_mean\_flux}, \texttt{phot\_rp\_mean\_flux}, and \texttt{phot\_bp\_mean\_flux} values of each source. 
The 2MASS XSC photometry is provided in AB magnitudes. For unWISE photometry, the conversion formula provided in \citet{unWISE2019} is applied to \textit{W1} and \textit{W2} \texttt{flux} values.
 
In addition to the photometric data mentioned above, we also add supplementary data.
\citet{MBRNN} used both the PSF and Kron magnitudes for each color index as inputs for their ML model; we adopt a similar strategy in this study. The photometric systems of the surveys considered here differ in their magnitude definitions; \textit{Gaia} magnitudes are measured using point/line spread function (PSF/LSF)  photometry \citep{GaiaEDR3Phot}, 2MASS XSC magnitudes are based on the $20 \;\mathrm{mag\;arcsec^{-2}}$ isophotal aperture \citep{2MASSXSC}, and unWISE magnitudes are derived from PSF fitting \citep{unWISE2019}. To incorporate point-source and extended-source photometry on an equivalent basis so that the structure of our input feature is set as similar as possible to that of \citet{MBRNN}, we use S\'ersic magnitudes provided in the \textit{Gaia} DR3 \texttt{galaxy\_candidates} table as the extended-source counterpart to the PSF-based magnitudes.
The S\'ersic magnitude is calculated by applying the conversion formula in Section 5.4.1 of the \textit{Gaia} DR3 \hyperlink{https://gea.esac.esa.int/archive/documentation/GDR3}{Online Documentation} that changes the flux to AB magnitude on the \texttt{intensity\_sersic} field of the \textit{Gaia} DR3 \texttt{galaxy\_candidates} table.
The \texttt{intensity\_sersic} is the intensity measured by \textit{Gaia} within the effective radius assuming the S\'ersic profile.
In addition, \texttt{sersic\_ellipticity} and \texttt{n\_sersic} fields are used to provide more information on the shape of the source.
Moreover, \texttt{in\_qso\_candidates} field is used to give information on whether a source is a potential QSO.
The \texttt{in\_qso\_candidates} values are set 1 if a source is included in the \textit{Gaia} DR3 \texttt{qso\_candidates} table, and 0 if not.
The details on the fields \texttt{intensity\_sersic}, \texttt{ellipticity\_sersic}, \texttt{n\_sersic}, and \texttt{in\_qso\_candidates} can be found in Section 20.5.1 of the \textit{Gaia} DR3 \hyperlink{https://gea.esac.esa.int/archive/documentation/GDR3}{Online Documentation}.
Finally, the $E(B-V)$ values, which are provided by \citet{Greiss2014} and can be accessed through \texttt{python} package \texttt{dustmaps} \citep{Green2018}, in the direction of each source are included.

The errors of each column are also included, except for \texttt{in\_qso\_candidates} which does not have associated uncertainties due to its boolean nature and $E(B-V)$ whose error is unobtainable by the \texttt{dustmaps} package.
The \texttt{NaN} values are replaced with $-9$.
The summary of the input data is shown in Table~\ref{InputTab}.

The subset of input data with the spec-$z$ is hereafter referred to as the spec-$z$ subset and that without spec-$z$ is referred to as inference subset.
The spec-$z$ subset is further divided randomly into three distinct subsets of 80\% training set, 10\% validation set, and 10\% test set.

\begin{table*}[!h]
    \centering
    \caption{Summary of input data}
    \begin{tabular}{c p{3.8cm} p{3.4cm} | c p{2.9cm} p{4.1cm}}
        \toprule
        No. & Column name & Description & No. & Column name & Description  \\ \hline
        1 & \texttt{G} & \textit{Gaia} $G$ mag & 13 & \texttt{in\_qso\_candidates} & if in \texttt{qso\_candidates} table \\
        2 & \texttt{G\_err} & \textit{Gaia} $G$ mag error & 14 & \texttt{j\_m\_k20fe0} & 2MASS \textit{J} mag \\
        3 & \texttt{rp} & \textit{Gaia} $G_{RP}$ mag & 15 & \texttt{j\_msig\_k20fe0} & 2MASS \textit{J} mag error \\
        4 & \texttt{rp\_err} & \textit{Gaia} $G_{RP}$ mag error & 16 & \texttt{h\_m\_k20fe0} & 2MASS \textit{H} mag \\
        5 & \texttt{bp} & \textit{Gaia} $G_{BP}$ mag & 17 & \texttt{h\_msig\_k20fe0} & 2MASS \textit{H} mag error \\
        6 & \texttt{bp\_err} & \textit{Gaia} $G_{BP}$ mag error & 18 & \texttt{k\_m\_k20fe0} & 2MASS \textit{K} mag \\
        7 & \texttt{sersic\_mag} & S\'ersic fit mag & 19 & \texttt{k\_msig\_k20fe0} & 2MASS \textit{K} mag error \\
        8 & \texttt{sersic\_mag\_err} & S\'ersic fit mag error & 20 & \texttt{w1mag} & unWISE \textit{W1} mag \\
        9 & \texttt{ellipticity\_sersic} & S\'ersic fit ellipticity & 21 & \texttt{dw1mag} & unWISE \textit{W1} mag error \\
        10 & \texttt{ellipticity\_sersic\_err} & S\'ersic fit ellipticity error & 22 & \texttt{w2mag} & unWISE \textit{W2} mag \\
        11 & \texttt{n\_sersic} & S\'ersic index & 23 & \texttt{dw2mag} & unWISE \textit{W2} mag error \\
        12 & \texttt{n\_sersic\_err} & S\'ersic index error & 24 & \texttt{E(B\!-\!V)} & Galactic Color Excess \\
        \bottomrule
    \end{tabular}
    \begin{tablenotes}
        \item \textbf{Note.} \texttt{NaN} values replaced with $-9$.
    \end{tablenotes}
    \label{InputTab}
\end{table*}

\begin{figure}[!h]
    \centering
    \includegraphics[width=0.4\textwidth]{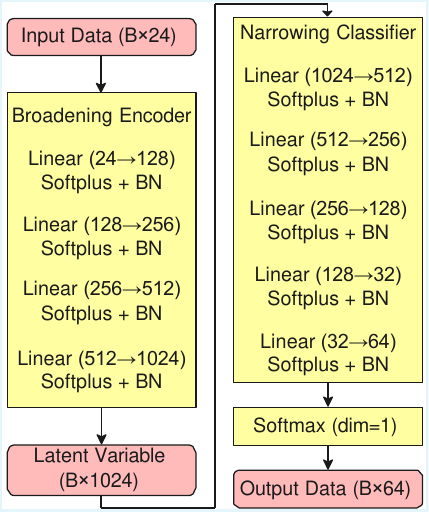}
    \caption{Structure of MBRNN model algorithm.}
    \label{MBRNNStructure}
\end{figure}

\section{Machine Learning Model: MBRNN}\label{SSMBRNN}
Multiple-Bin Regression with Neural Network (MBRNN) is a neural network-based multi-layer perceptron (MLP) algorithm that predicts the probability of the photo-$z$ of an object lying in each element of a predefined redshift binning scheme \citep{MBRNN}.
The binning scheme may be set however the users want it.
In this study, it is defined so that there are 64 bins in total.
The first 51 are evenly spaced bins that span over $-0.082\leq z \leq 1$.
The rest have widths of 0.5, covering $1\leq z\leq7$. 

Despite the name, MBRNN is not a recurrent neural network (RNN) but rather a feed-forward architecture consisting of nine fully connected (FC) linear operation layers.
The first four layers expand the dimension of the vector from the number of input features (24 in this study) to 1024. 
The subsequent four layers progressively reduce the dimensionality to 32, and the final layer maps the vector so that the output matches the number of redshift bins (64 in this case).
Consecutive FC layers are connected via softplus activation functions, which is then followed by batch normalization (BN). 
The weights of the learnable parameters are initialized using the Xavier initialization method. \citep{GlorotBengio2010}
This structure is visualized in Figure \ref{MBRNNStructure}.

When the model is fed with the training and validation set, it starts to find the best parameters using a logistic regression.
The anchor loss function that the model was trained to minimize is given as
\begin{align*}
    \mathrm{Loss}(x, y) = \sum_k &- y_k \log\left( \max\lbrace x_k, 10^{-8}\rbrace \right) \\
    &- (1 - y_k) \log\left( \max\lbrace1 - x_k, 10^{-8}\rbrace \right),
\end{align*}
where $x_k$ is the predicted probability for the $k$-th bin, $y_k=1$ if the photo-$z$ is in the correct redshift bin (case referred to positive class) and $y=0$ otherwise (negative class).
Unlike in the original implementation of \citet{MBRNN}, a clipping value $10^{-8}$ is used within the logarithm to ensure numerical stability and avoid computing $\log(0)$.
Furthermore, the original paper generalized this logistic regression by applying the exponent factor $\gamma$, which is adjustable by the user, to term $(1-y_k)$, and it is set to 0 in this study.
That is, the ordinary logistic regression is used.

The Adam method \citep{Adam} is adopted for the optimization method.
The parameter $\beta_1$, which governs the momentum of the descent, is set to be 0.5, and $\beta_2$, which controls step sizes according to the $L^2$ norm of the gradient, is set to be 0.999.
The gradient is clipped so that it does not exceed 5.
The weight of the $L^2$ regularization term in the objective function is set to $5\times10^{-5}$.
The initial learning rate is set to 0.0008, and scheduled to halve when the validation loss does not improve for five consecutive epochs, until the learning rate becomes 0.000001.

The machine yields the probability distribution of photo-$z$ lying in each bin as the output.
The probability-weighted average is considered to be the photo-$z$, and the standard deviation of the probability distribution is used as the error of the photo-$z$ ($\sigma_{z_{\mathrm{phot}}}$). 

\citet{MBRNN} explored multiple ML-based photo-$z$ estimation models: $k$-nearest neighbors, random forest, vanilla neural network, and MBRNN.
They trained galaxies in the Pan-STARRS1 catalog using these four models and found that the MBRNN model showed the best performance.
Consequently, this study also adopts the MBRNN model based on its demonstrated superior performance.

\section{Results}
\subsection{Spectroscopic Redshifts}
\begin{figure}[!h]
    \centering
    \includegraphics[width=0.5\textwidth]{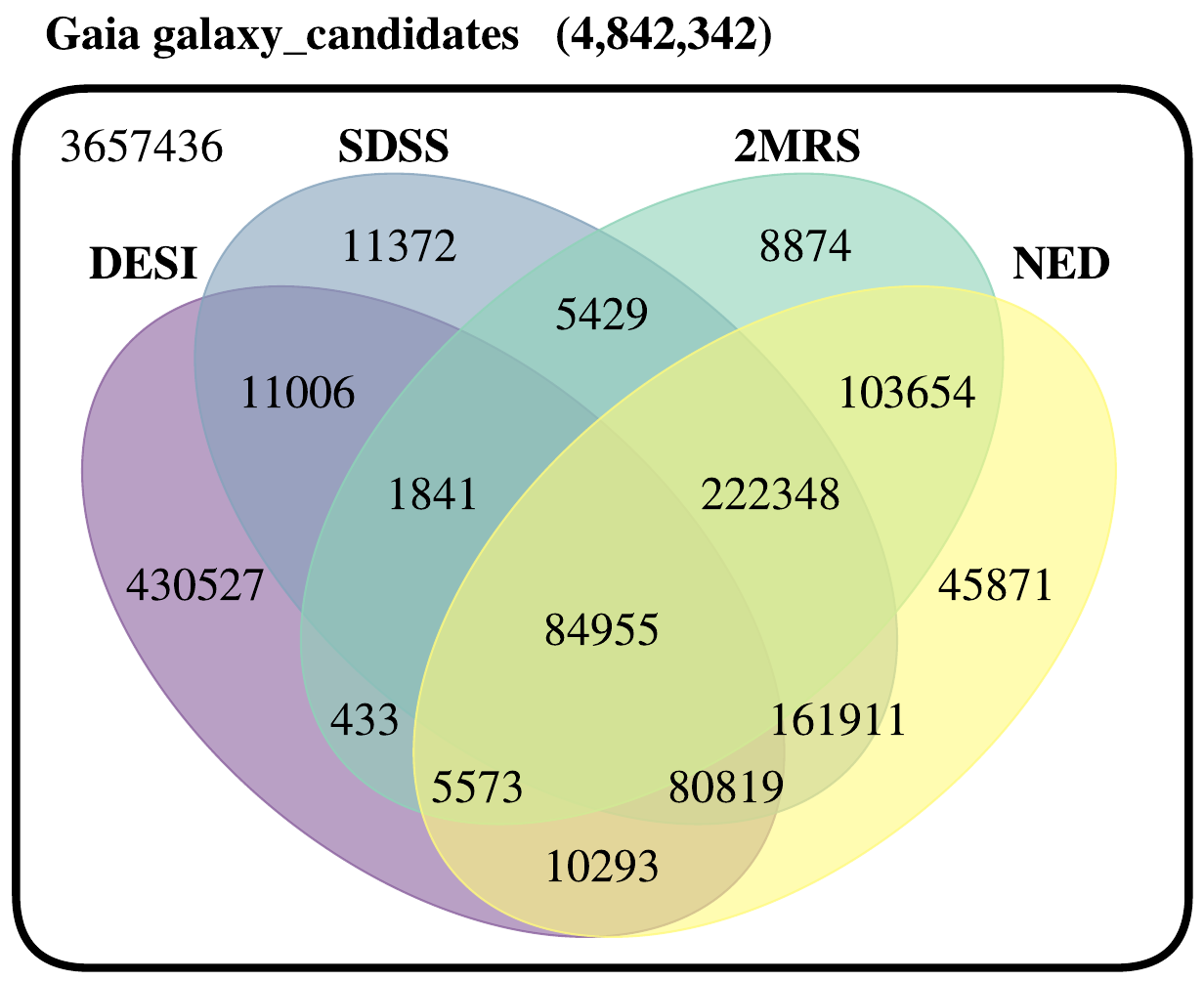}
    \caption{Venn diagram of \texttt{galaxy\_candidates} table and its subsets matched to the spec-$z$ catalogs}
    \label{SpecVenn}
\end{figure}

\begin{figure*}[!h]
    \centering
    % First Image
    \includegraphics[width=0.48\textwidth]{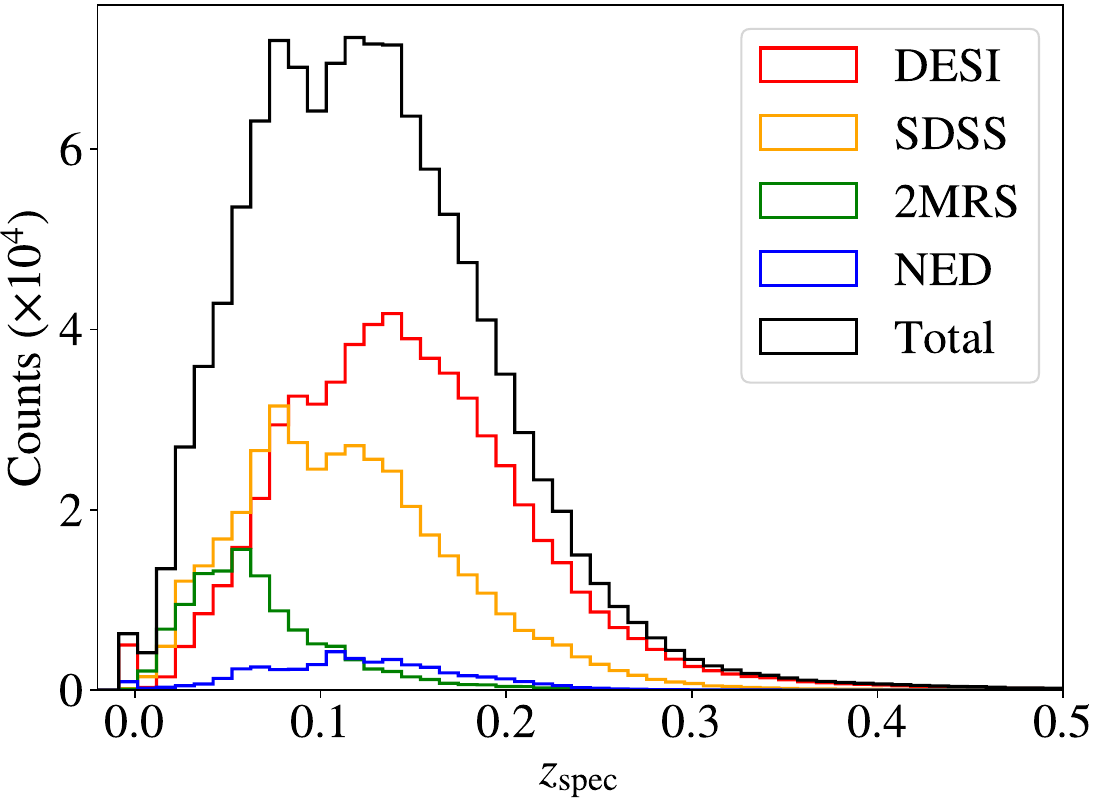}
    % Second Image
    \includegraphics[width=0.48\textwidth]{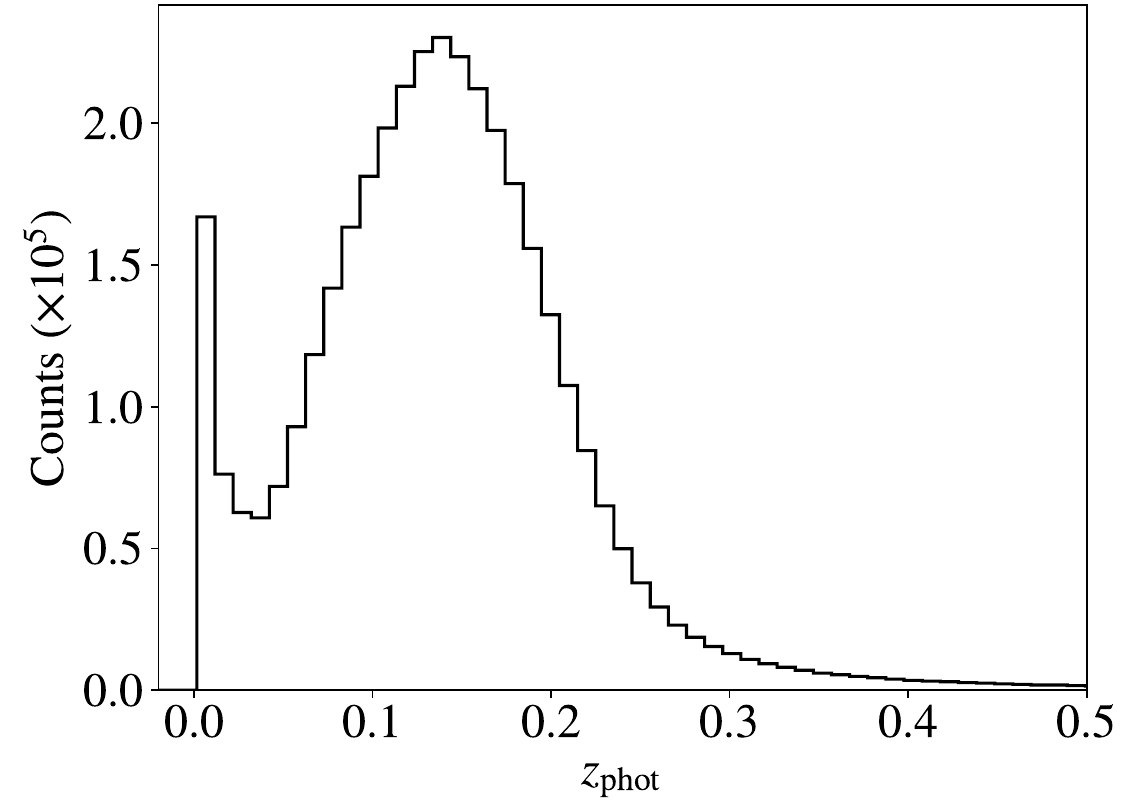}
    
    \caption{The histrogram of spectroscopic redshifts matched to \textit{Gaia} DR3 \texttt{galaxy\_candidates} (left panel), and the histogram of the estimated photometric redshifts of inference set (right panel). For clarity, the redshift ranges are limited up to 0.5, even though there are sources up to $z_\mathrm{spec}=6.88$ and $z_\mathrm{phot}=2.82$.}
    \label{Hist}
\end{figure*}

Among 4.8 million sources in the \textit{Gaia} DR3 \texttt{galaxy\_candidates} table, 1,184,906 sources (24.5\%) have counterparts in at least one of the four spec-$z$ catalogs.
A Venn diagram summarizing the results of cross-matching is presented in Figure~\ref{SpecVenn}.
The histogram of the spec-$z$ of the matched \texttt{galaxy\_candidates} sources according to this law is shown in the left panel of Figure~\ref{Hist}.

\subsection{Photometric Redshifts}
\begin{figure}[!h]
    \centering
    \includegraphics[width=0.5\textwidth]{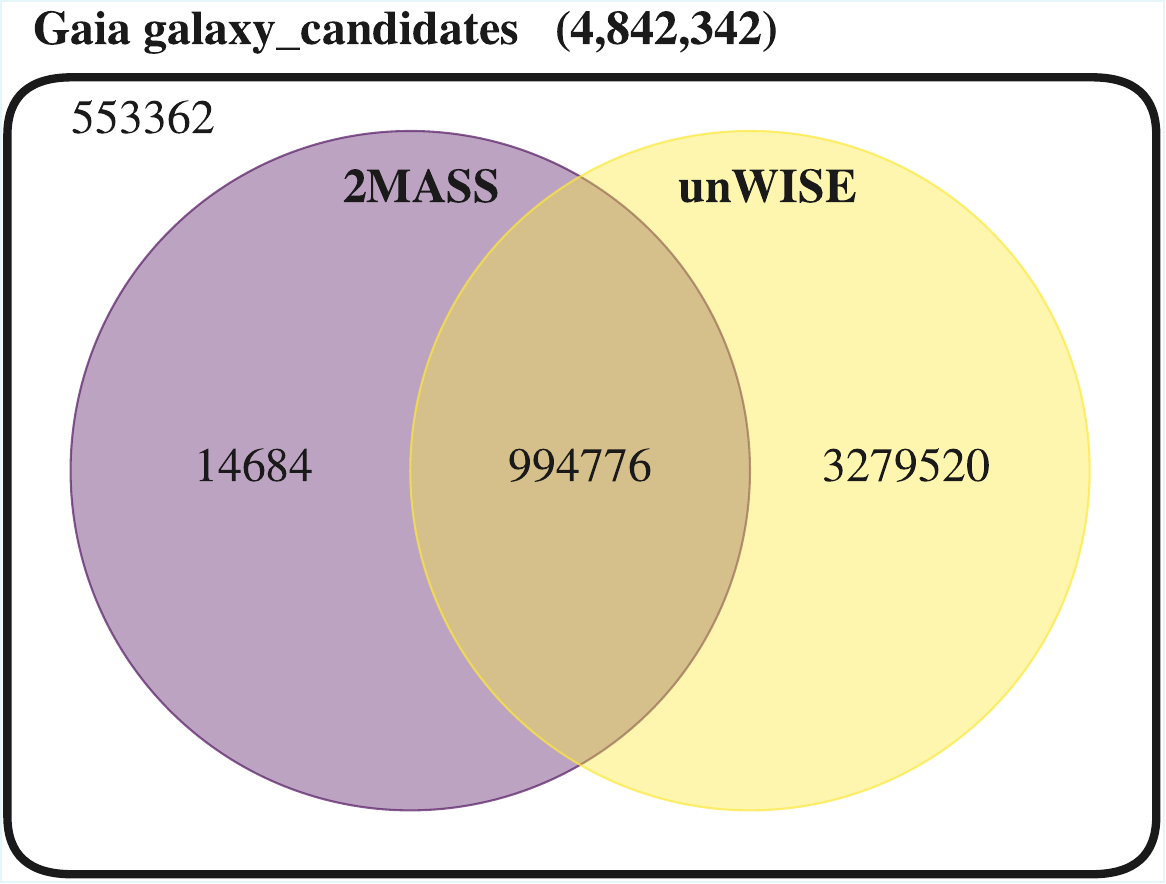}
    \caption{Venn diagram of \texttt{galaxy\_candidates} table and its subsets matched to 2MASS and unWISE catalogs}
    \label{PhotFig}
\end{figure}

A Venn diagram summarizing the results of photometric catalog cross-matching is presented in Figure~\ref{PhotFig}.

\begin{figure*}[!h]
    \centering
    \includegraphics[width=\textwidth]{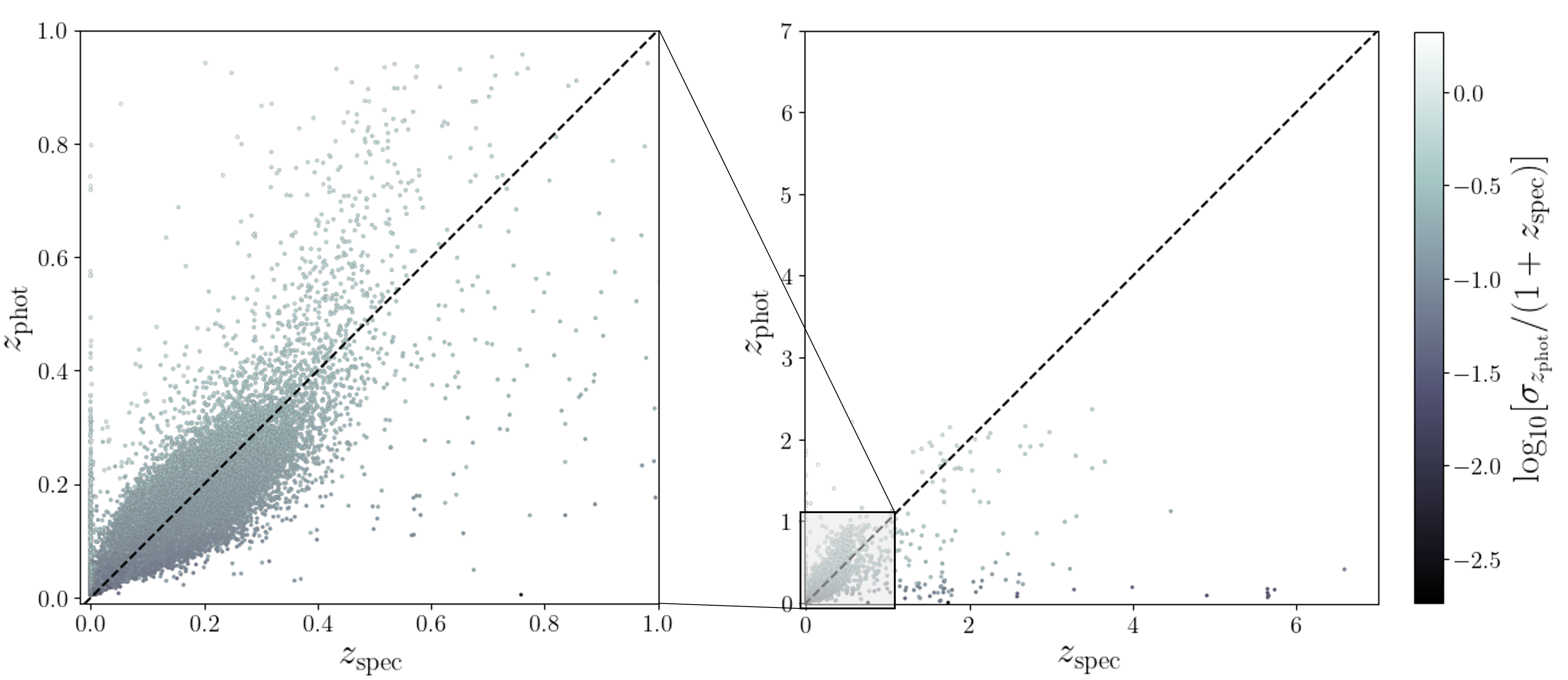}
    \caption{Photo-$z$ and spec-$z$ plot for the test set. The color indicates the error of the photo-$z$.}
    \label{ResultPlot}
\end{figure*}

In Figure \ref{ResultPlot}, the photo-$z$ inference result of the test set is shown and its spec-$z$ is shown.
The color indicates the error of the photo-$z$.
The statistical property of this plot is shown in Table \ref{tab:ResultStat}.
In this table, $\Delta z$ is defined as the residual distribution $z_{\mathrm{phot}} - z_{\mathrm{spec}}$. 
The statistical metrics are defined as follows:

\begin{itemize}
    \item {Bias}: The absolute value of the mean of $\Delta z / (1 + z_{\mathrm{spec}})$.
    \item {MAD (Mean Absolute Difference)}: The mean value of $|\Delta z / (1 + z_{\mathrm{spec}})|$.
    \item {$\sigma$}: The standard deviation of $\Delta z$.
    \item {$\sigma_{68}$}: The 68th percentile of $|\Delta z|$.
    \item {NMAD (Normalized Median Absolute Deviation)}: $1.4826 \times \text{median}(|\Delta z|)$.
    \item {$R_{\mathrm{cat}}$}: The fraction of sources where $|\Delta z / (1 + z_{\mathrm{spec}})| > 0.15$.
\end{itemize}
These metrics are used by \citet{Cavuoti2017} and \citet{Salvato2019} as well as \citet{MBRNN} to test the accuracy and precision of different ML models.
The histogram of the inferred photo-$z$ of the inference set is shown in the right panel of Figure~\ref{Hist}.

\begin{table*}[!h]
    \centering
    \caption{Metric analysis of the test set.}
    \begin{tabular}{cccc}
        \toprule
        Metric & Description & Value (24-feat) & Value (15-feat) \\ \hline
        Bias & $\left|\mathbb{E}\left[\Delta z / (1+z_\mathrm{spec})\right]\right|$ & 0.0014 & 0.0032 \\
        MAD & $\mathbb{E}\left[\left|\Delta z / (1+z_\mathrm{spec})\right|\right]$ & 0.0192 & 0.0195 \\
        $\sigma$ & $\sigma(\Delta z)$ & 0.0739 & 0.0775 \\
        $\sigma_{68}$ & $68^{\mathrm{th}}$ percentile $\left|\Delta z\right|$ & 0.0228 & 0.0231\\
        NMAD & $1.4826\times\mathrm{Median}\left(\Delta z / (1+z_\mathrm{spec})\right)$ & 0.0021 & 0.0021\\
        $R_\mathrm{cat}$ & $\left|\Delta z / (1+z_\mathrm{spec})\right|>0.15$ fraction & 0.0048 & 0.0051 \\
        \bottomrule
    \end{tabular}
    \begin{tablenotes}
        \item Note. $\Delta z = z_\mathrm{phot}-z_\mathrm{spec}$
    \end{tablenotes}
    \label{tab:ResultStat}
\end{table*}

\begin{figure*}[!h]
    \centering
    \includegraphics[width=0.9\linewidth]{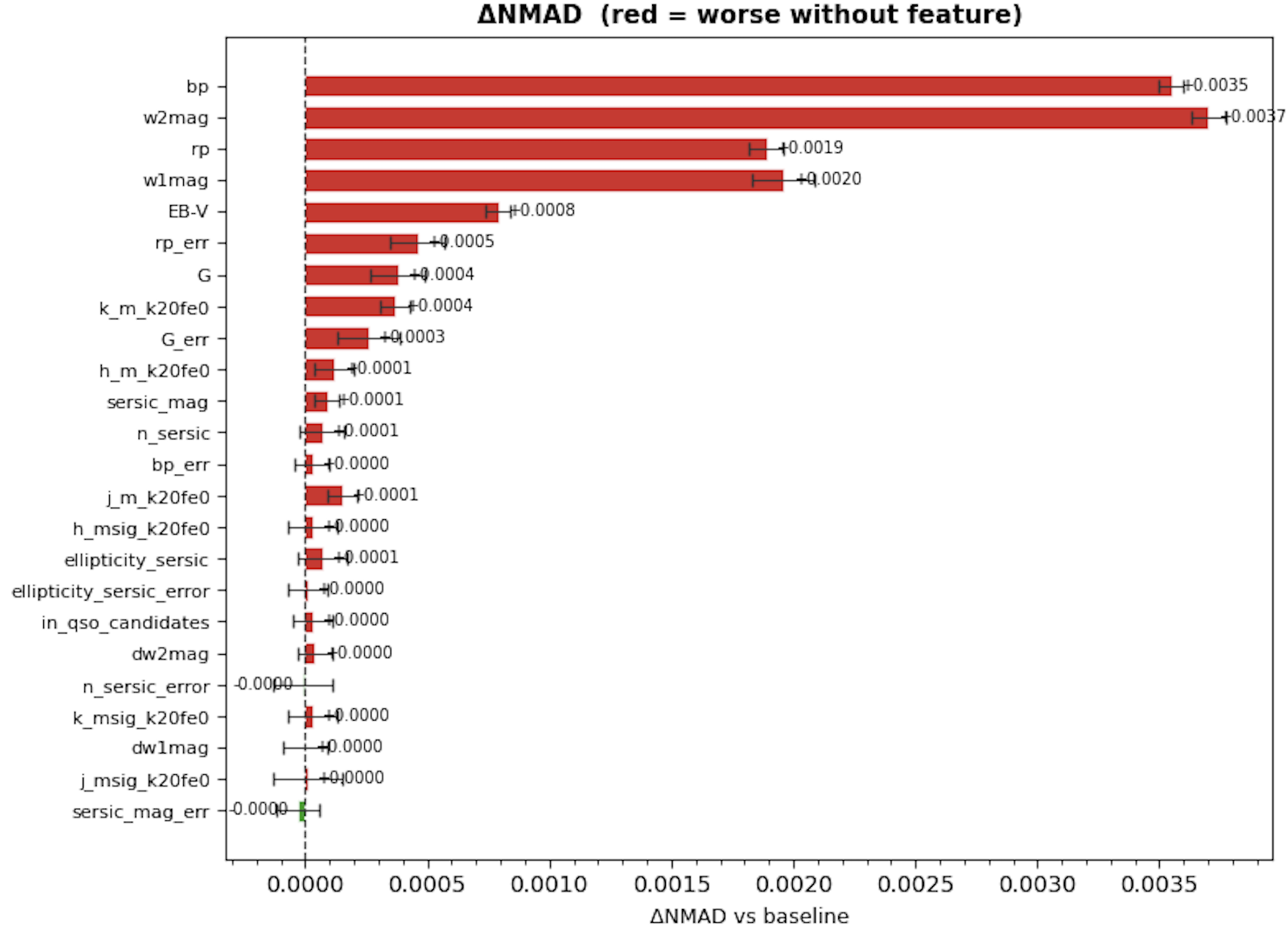}
    \caption{Ablation analysis based on $\Delta$NMAD. Red bars indicate that the
    model performed worse without that feature, while green bars indicate
    that it performed marginally better. No individual feature showed a
    significant improvement in performance upon removal, insufficient to
    justify exclusion.}
    \label{fig:ablation_comparison}
\end{figure*}

To assess the contribution of each input feature, we perform an ablation analysis in which each of the 24 features is removed individually and the MBRNN model is retrained on the reduced feature set. 
To account for variance in model performance, each ablation experiment is repeated five times using 5-fold cross-validation on the training set. 
The resulting variation across folds provides an estimate of the uncertainty in each metric, allowing us to determine whether the performance change upon feature removal is statistically significant. Figure~\ref{fig:ablation_comparison} shows the result of this analysis. The horizontal axis corresponds to the NMAD using 23 features minus the NMAD using all 24 features, i.e., the change in NMAD upon removing a given feature. NMAD is adopted as the primary evaluation metric, as it provides a robust estimate of the scatter in photometric redshift residuals that is insensitive to catastrophic outliers \citep{Brammer2008}. The remaining five metrics ($\Delta$Bias, $\Delta$MAD, $\Delta\sigma$, $\Delta\sigma_{68}$, and $\Delta R_{\rm cat}$) yield consistent results, with no feature showing statistically significant grounds for exclusion.

To further test this result, we retrain the model using the same training-validation-test split, but with only 15 input features, excluding the 9 features (errors in morphology, errors in IR bands, and the QSO candidate flag) that turn out to make only an insignificant contribution. 
The result of the metric analysis applied to this reduced model is given in the right-most column of Table~\ref{tab:ResultStat}. 
All six metrics derived using these 15 features are larger than or equal to those derived using all 24 input features, confirming that no subset of features improves model performance. 
Therefore, in the remaining sections, we discuss the results obtained using all 24 input features.

\begin{figure}[!h]
    \centering
    \includegraphics[width=0.5\textwidth]{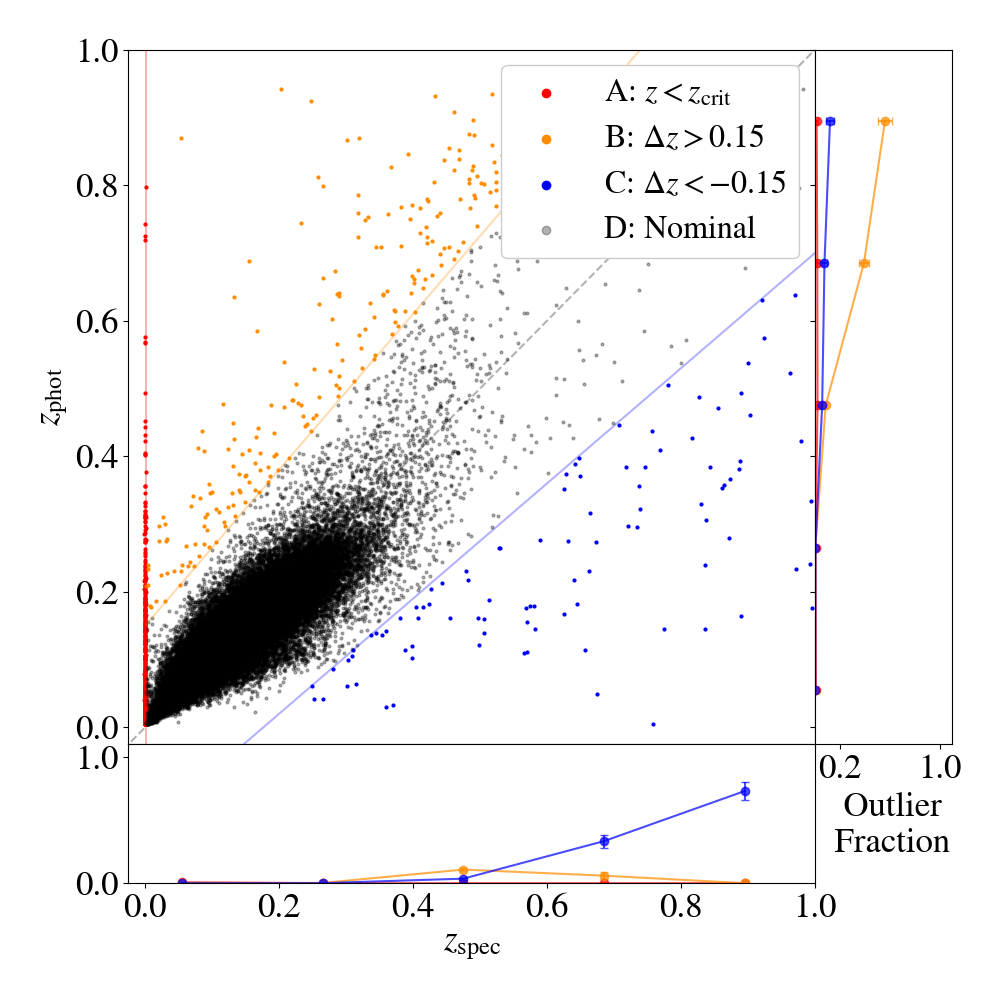}
    \caption{Plot of photometric redshifts and their spectroscopic redshifts of the test set sources. The set A, which is indicated by red, corresponds to $z_\mathrm{spec}<z_\mathrm{crit}\approx0.0017$, set B in orange to $\Delta z / (1+z_\mathrm{spec})>0.15$, set C in blue to $\Delta z / (1+z_\mathrm{spec})<-0.15$, set D in black (grey) to the nominal case that does not fall into any one of A, B and C. The solid lines indicate the boundaries of the outlier subsets, and the dashed line correspond to $z_\mathrm{spec}= z_\mathrm{phot}$. For a better visual, the redshift ranges are limited up to 1.}
    \label{DiscussionPlot}
\end{figure}

\section{Discussion}\label{discuss}
\subsection{Performance on Photometric Redshift Estimation}
All metrics in Table \ref{tab:ResultStat}, except for $\sigma$, are smaller than those reported in the best performance result obtained by \citet{MBRNN}. 
The higher value of $\sigma$, almost twice as high, is mainly due to the wide-extended redshift range to $z\approx7$.
This is evident from the right panel of Figure \ref{ResultPlot}, and when limited to $z=1$, $\sigma$ also becomes smaller than in the previous work.

The distribution of the photo-$z$ inference (the right panel of Figure \ref{Hist}) is similar to that of the input spec-$z$ (the left panel of Figure \ref{Hist}) for $z\gtrsim0.04$: having peaks in $0.13\lesssim z\lesssim0.14$, and rapidly decreasing for larger redshifts.
In particular, both histograms decrease below 10\% of the maximum count for bins in $z\gtrsim0.3$.
However, there are a couple of noticeable differences.
First, the inference set exhibits a prominent peak near $z\sim0$, which is absent in the input spec-$z$ distribution.
This discrepancy may be due to a higher fraction of stellar contaminants in the parent photometric sample compared to the spec-$z$ training samples or reflect the limitation of the machine learning model.
The exact cause remains uncertain.
Secondly, the input spec-$z$ distribution included negative redshifts, whereas the photo-$z$ inference had the lower bound of 0.005.
The reason for this discrepancy is unclear, but could again be related to limitations of the machine learning model, such as its inability to extrapolate beyond the range of well-represented training examples.

\subsection{Outlier Analysis}
Although the model demonstrated reasonable performance, as discussed in the previous section, there are still sources that clearly fall into the category of outliers. 
Analyzing these cases can offer valuable insight, allowing the assignment of a score $P_0$ that reflects the risk of either misidentifying a source as extragalactic or trusting an unreliable photometric redshift, by investigating its photometric properties.

The test result is categorized into four distinct subsets: sources with $z_\mathrm{spec}<z_\mathrm{crit}\coloneqq500 \;\mathrm{km\;s^{-1}}/c\approx0.0017$ (hereafter the set A), sources with $\Delta z / (1+z_\mathrm{spec})$ values greater than 0.15 (the set B), sources with $\Delta z / (1+z_\mathrm{spec})$ values smaller than $-0.15$ (the set C), and nominal cases that do not fall into any of the previous three subsets (the set D).
The velocity threshold of $500 \;\mathrm{km\;s^{-1}}$ is chosen as it roughly corresponds to the maximum escape velocity of the Milky Way \citep{Roche2024}, which implies that set A consists primarily of Galactic objects.
The numbers of sources in the outlier subsets A, B, and C are 637 $(R_\mathrm{A}=0.83\%)$, 236 $(R_\mathrm{B}=0.31\%)$, and 200 $(R_\mathrm{C}=0.26\%)$, respectively.
$R_\mathrm{outlier}$ refers to the fraction of a certain set of outliers.
This classification is illustrated in Figure \ref{DiscussionPlot}.

\begin{figure*}[!h]
    \centering
    \includegraphics[width=\textwidth]{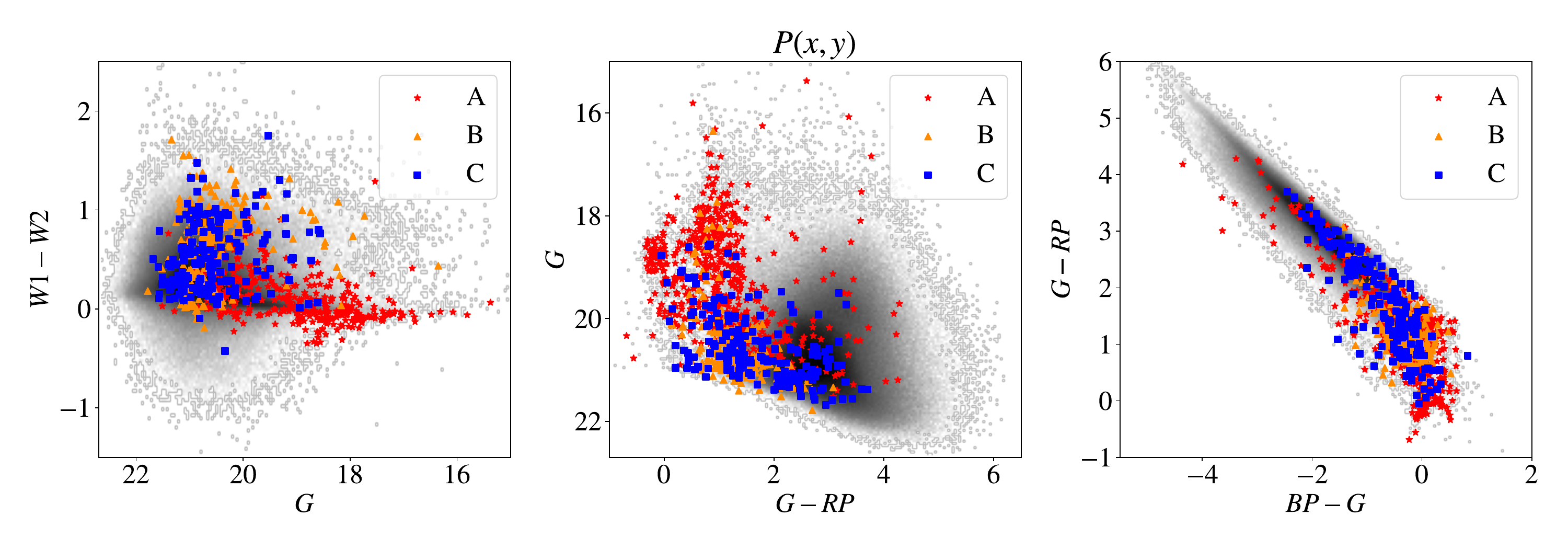}
    \caption{Plot of the outliers in three different color(magnitude) spaces: (\textit{G}, \textit{W1}$-$\textit{W2}) (left), (\textit{G}$-$\textit{RP}, \textit{G}) (middle), (\textit{BP}$-$\textit{G},\textit{G}$-$\textit{RP}) (right). Sources in set A are indicated as red star-shaped points, set B sources with orange triangles and set C sources with blue squares. The grayscale plots in the background are the evidence distributions of spec-$z$ subset sources.}
    \label{VarSpace}
\end{figure*}

\begin{figure*}[!h]
    \centering
    \includegraphics[width=0.9\textwidth]{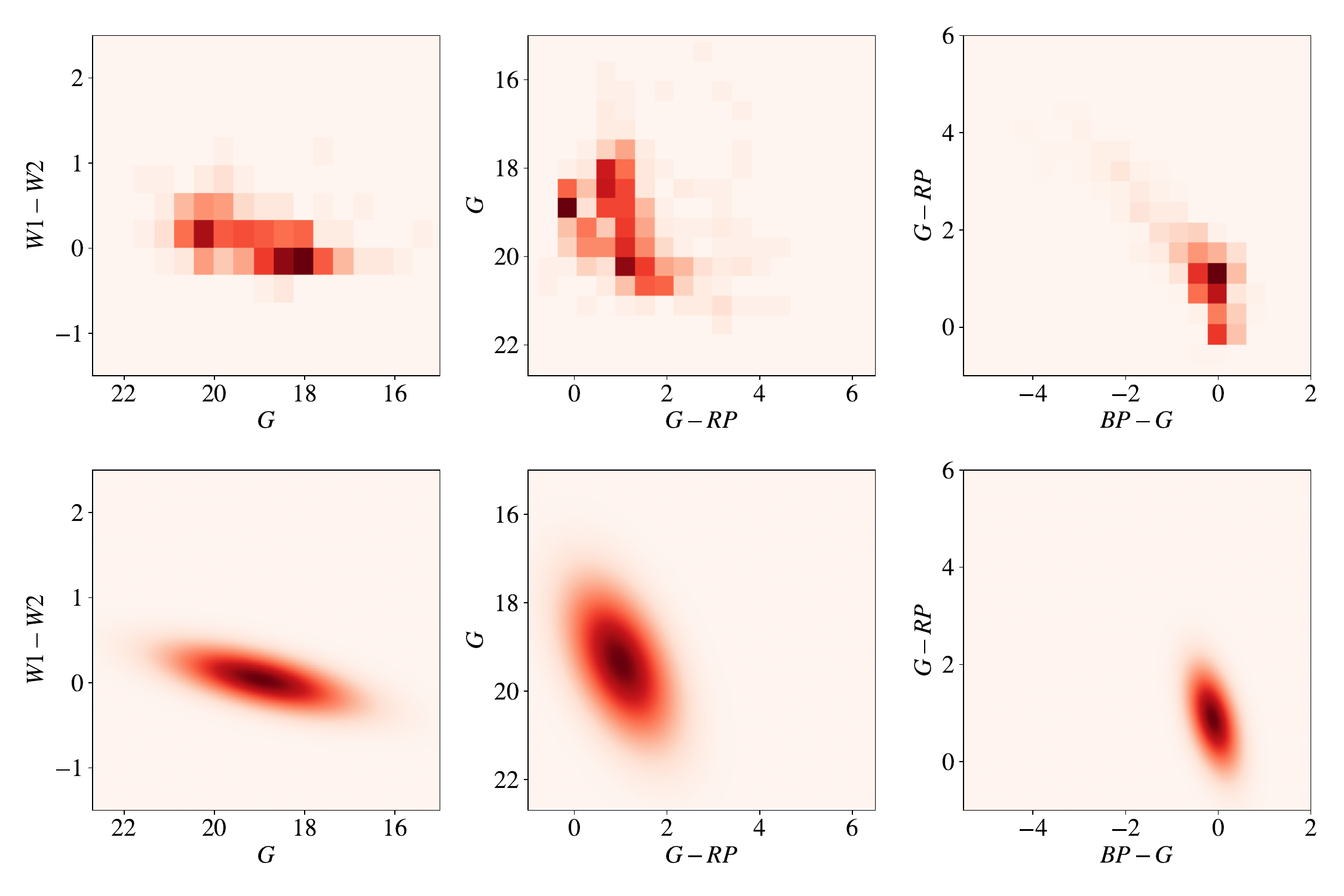}
    \caption{Plot of the likelihood that a set A source being located at certain points for three different color (magnitude) spaces (upper row), and their Gaussian Fitting. (Lower row) Similar processes are conducted to all of the three outlier sets.}
    \label{RedFit}
\end{figure*}

To construct a posterior probability distribution function of a source falling into one of the outlier categories given a set of photometric data, the outliers are first plotted in three different color-color or color-magnitude spaces: (\textit{G}, \textit{W1}$-$\textit{W2}), (\textit{G}$-$\textit{RP}, \textit{G}) and (\textit{BP}$-$\textit{G},\textit{G}$-$\textit{RP}).
At the same time, the distributions of all sources in the spec-$z$ subset are also plotted in these three spaces as two-dimensional histograms for comparison.
The space (\textit{G}, \textit{W1}$-$\textit{W2}) is known to be effective in distinguishing QSOs from galaxies \citep{GaiaAGN2022}.
Similarly, the spaces (\textit{G}$-$\textit{RP}, \textit{G}) and (\textit{BP}$-$\textit{G},\textit{G}$-$\textit{RP}) have proven useful in separating stars and QSOs from galaxies \citep{GaiaCollaboration2023}.

In Figure \ref{VarSpace}, each of the outlier sets is plotted in the three color-color (color-magnitude) spaces, along with the probability distribution of the spec-$z$ subset.
The plot implies that set A consists mainly of stars and sets B and C consists of galaxies and QSOs that are dimmer than $G\lesssim20^\mathrm{mag}$.
The probability of being in set A is higher in the regions $G\lesssim 2^\mathrm{mag}$ of the $(G-RP, G)$ space and $(BP-G)-(G-RP)\lesssim3^\mathrm{mag}$ of the $(BP-G, G-RP)$ space.
According to \citet{GaiaCollaboration2023}, this is where the stars and QSOs usually distribute.
Since the radial velocities of the sources in the red set are smaller than the escape velocity of the Milky Way, these sources are probably stars.
Some sources in sets B and C show similar properties in the $(G-RP, G)$ and $(BP-G, G-RP)$, and others spread into the $G-RP\gtrsim 2$ region, where galaxies are.
Moreover, in $(G, W1-W2)$ space, some of these objects have $W1-W2>0.8$, where QSOs are, and some have smaller WISE color, which means that these are either stars or galaxies \citep{Wright2010, Carnerero2023}.
As these objects have large spec-$z$, they are likely galaxies.
They are also usually fainter than 20 magnitude in the G band, whereas red set stars are mostly brighter.

Next, each outlier subset, identified by color in Figure~\ref{VarSpace}, is converted to a low-resolution 2d histogram, as shown in the top row of Figure~\ref{RedFit}.
The resolution of the histograms is then enhanced by fitting the distribution to two-dimensional Gaussian functions.
To match the resolution of the spec-$z$ subset distribution histograms, these outlier histograms are refined by fitting two-dimensional Gaussian functions:
$$f(\mathbf{x})\approx \frac{1}{(2\pi)^{d/2}|\mathbf{\Sigma}|^{1/2}}\exp\left[-\frac{1}{2}(\mathbf{x}-\mathbf{\mu})^\mathbf{T}\mathbf{\Sigma}^{-1}(\mathbf{x}-\mathbf{\mu})\right].$$
Here $\mathbf{\mu}, \mathbf{x}\in \mathbb{R}^d$, $\mathbf{\Sigma}\in\mathbb{R}^{d\times d}$, with $d=2$ and $\mathbf{x}$ representing the coordinates in one of the three color-color (color-magnitude) spaces.
There are two fitting parameters in $\mu$, and three independent fitting parameters in $\Sigma$.
The resulting fitted distributions for the red outlier set are presented in the lower row of Figure \ref{RedFit}.

\begin{figure*}[!h]
    \centering
    \includegraphics[width=\textwidth]{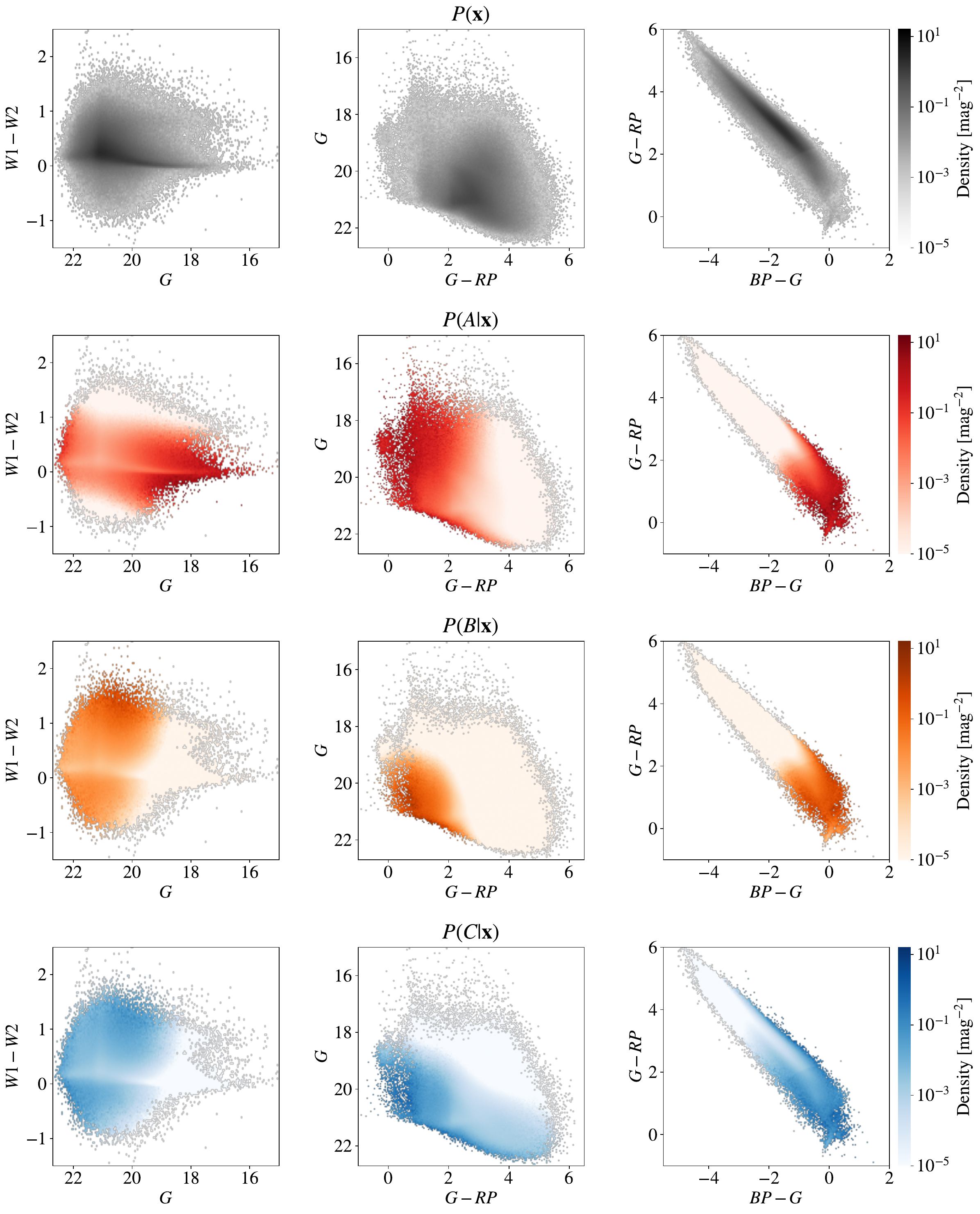}
    \caption{Probability in the spec-$z$ catalog being found in a certain location in different spaces. (Top row) Conditional probability plot of an object at a given location falling into a catastrophic case. (2, 3, and 4th rows. Each correspond to the red, orange, and blue catastrophic cases, respectively.) Units are in mag$^{-2}$}
    \label{ProbScore}
\end{figure*}

Note that fitted distributions are equivalent to the likelihood probability distributions in color spaces given a specific outlier classification type, that is, $f(\mathbf{x})=P(\mathbf{x}|\mathrm{outlier})$.
On the other hand, the reliability score refers to the posterior probability that a source belongs to an outlier class given its photometric properties, that is, $P(\mathrm{outlier}|\mathbf{x})$.
These two quantities are conditional probabilities with the roles of the event and the condition reversed, which are related through Bayes' theorem:
$$P(\mathrm{outlier}|\mathbf{x}) = \frac{P(\mathbf{x}|\mathrm{outlier})P(\mathrm{outlier})}{P(\mathbf{x})}$$
Here, $\mathrm{outlier}\in \lbrace\mathrm{A},\mathrm{B},\mathrm{C}\rbrace$.
Moreover, the prior distribution is $P(\mathrm{outlier})=R_\mathrm{outlier}$ and the evidence distribution $P(\mathbf{x})$ is the  distribution of spec-$z$ subset.
The evidence distribution $P(\mathbf{x})$ is shown in the uppermost row of Figure \ref{ProbScore} and the posterior distribution $P(\mathrm{outlier}|\mathbf{x})$ for each outlier in sets A, B and C are shown in the lower three rows, respectively.

\begin{figure}[!h]
    \centering
    \includegraphics[width=0.5\textwidth]{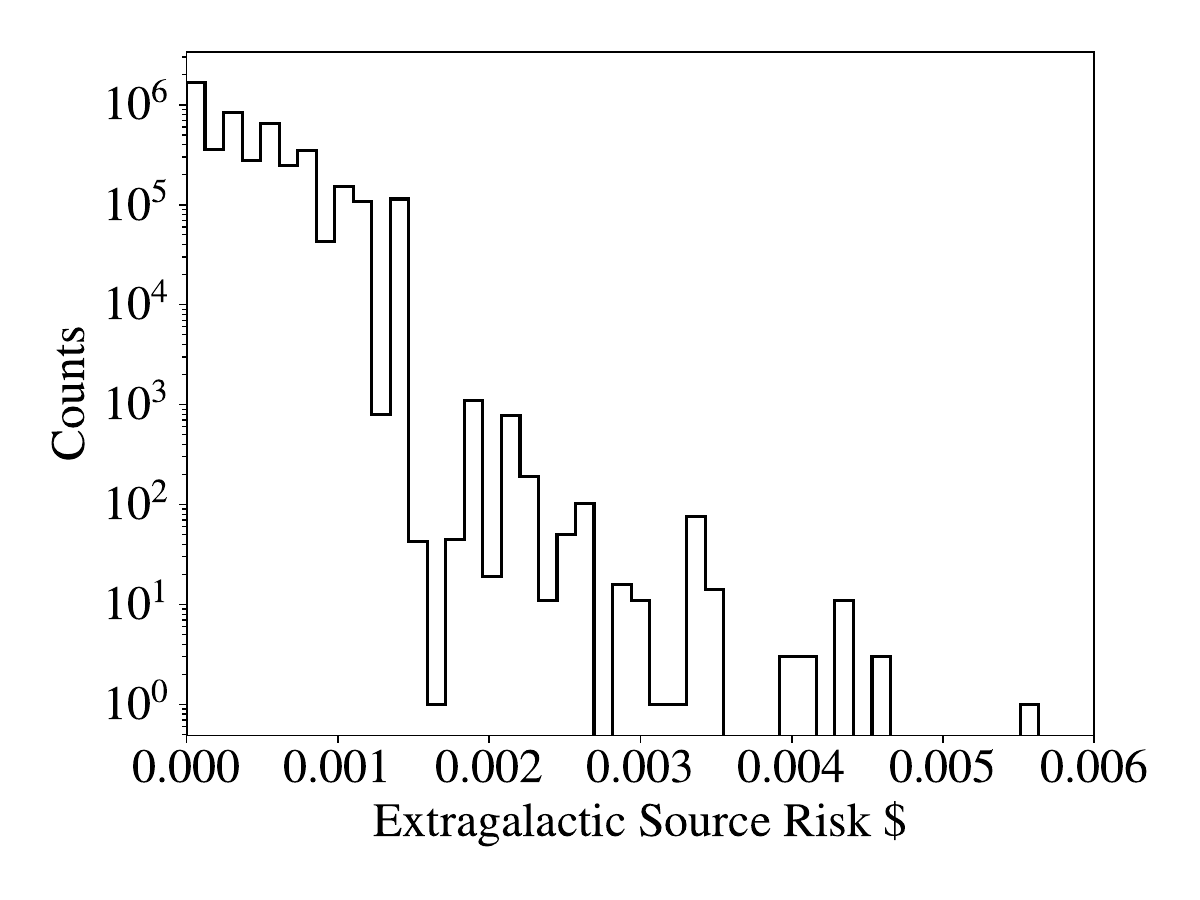}
    \caption{Histogram of $P_0$, a combined measure of the the risk of trusting it as a genuine extragalactic source or trusting its photometric redshift. The smaller $P_0$ is, the more reliable that source is as a galaxy.}
    \label{OutHist}
\end{figure}

$P_0$ is defined as the maximum of $P(\mathrm{A}|\mathbf{x})$, $P(\mathrm{B}|\mathbf{x})$, and $P(\mathrm{C}|\mathbf{x})$ for the sources in the inference subset, serving as a combined measure of the risk that a source is Galactic and the risk that its photometric redshift is catastrophically wrong. 
Here, set A consists primarily of stars, and sets B and C are the subsets that lie far from the $z_\mathrm{spec} = z_\mathrm{phot}$ line. 
Sources in subset A are assigned $P_0 = 1$, since their spectroscopic redshifts fall below that corresponding to the escape velocity of the Milky Way, suggesting they are Galactic contaminants rather than genuine extragalactic sources. 
Sources in subsets B, C, and D are assigned $P_0 = 0$, as their redshifts exceed this threshold and they are therefore almost certainly extragalactic. The smaller $P_0$ is, the more reliable that source is as a galaxy. 
The distribution of $P_0$ is shown in Figure \ref{OutHist}.

Finally, we construct an enhanced catalog based on the \textit{Gaia} DR3 \texttt{galaxy\_candidates} table, which includes spec-$z$ and photo-$z$, a redshift flag (\texttt{z\_flag}), and the outlier score.
The redshift flag is set to 0 for spectroscopic redshifts, 1 for photometric redshifts, and 2 when $P_0$ could not be computed due to insufficient photometric data.
There are 1,184,906 sources with \texttt{z\_flag}=0 (24.5\%), 3,601,132 sources with \texttt{z\_flag}=1 (74.4\%), and 56,304 sources with \texttt{z\_flag}=2 (1.2\%). 
Sources with a redshift flag equal to 2 always have photometric redshifts and no spectroscopic redshift, and are set to be $P_0=-9$.
A few rows of the finalized redshift catalog are presented in Table \ref{GaiaRedshiftCatalog}.

\begin{figure*}[!h]
    \centering
    \includegraphics[width=\textwidth]{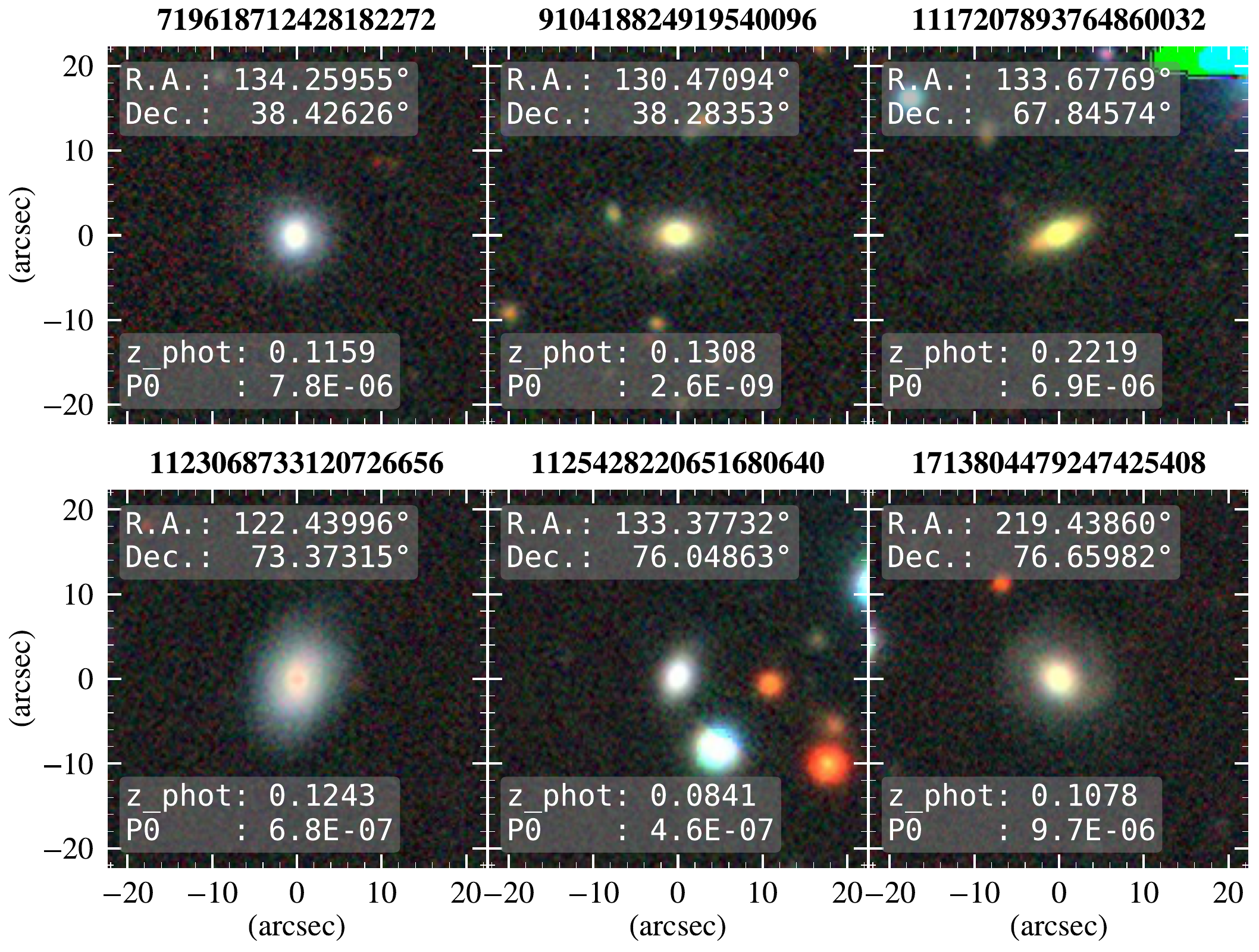}
    \caption{Multi-panel postage stamps of five newly discovered galaxies. Each panel presents a $45'' \times 45''$ ($0.75' \times 0.75'$) cutout from the Legacy Surveys DR10. Titles correspond to the \textit{Gaia} DR3 \texttt{SOURCE\_ID}. Equatorial coordinates are overlaid at the top of each panel, while photometric redshifts ($z_{\mathrm{phot}}$) and outlier scores ($P_0$) are provided at the bottom.}
    \label{NewGals}
\end{figure*}

Investigating sources with small $P_0$ values, we were able to discover some new galaxies.
These include sources indicated in Figure \ref{NewGals}.
The Legacy Surveys DR10 $grz$-color composite images \citep{Dey2019}, which are available in the Legacy Survey Sky Viewer\footnote{\url{https://www.legacysurvey.org/viewer}}, clearly show images of galaxies, as they are more extended compared to nearby stars.
These sources had outlier score values of $P_0\lesssim10^{-5}$, which is relatively small, according to Figure \ref{OutHist}.

\section{Summary}
This study enhances the \textit{Gaia} DR3 \texttt{galaxy\_candidates} table by incorporating both spec-$z$ and photo-$z$ measurements. 
We first aggregated the table with spec-$z$ values obtained via cross-matching with DESI DR1, SDSS DR17, 2MRS, and NED query result. 
For the remaining sources, photo-$z$ estimates were derived using the MBRNN model, which was trained on these spectroscopic datasets in conjunction with 2MASS and unWISE photometry.

Furthermore, we introduced a probabilistic outlier classification system based on color-magnitude and color-color spaces.
This allowed identification and characterization of outliers, including Milky Way contaminants (which were tagged as set A) and major photo-$z$ failures (which were tagged as sets B and), and the assignment of reliability scores to individual sources using a Bayesian framework.
In fact, it was possible to find new galaxies with the help of the probability score.

These improvements significantly enhance the scientific utility of the \textit{Gaia} extragalactic catalog, making it a more robust resource for statistical and cosmological studies. 
The newly constructed data set enables scientific exploration of the universe while providing built-in quality metrics to support informed source selection.

%%% ACKNOWLEDGMENTS (IF ANY) %%%%%%%%%%%%%%%%%%%%%%%%%%%%%%%%%%%%%%%%

\acknowledgments
We thank the reviewer for the insightful comments. HSH acknowledges support from the National Research Foundation of Korea (NRF) funded by the Korea government (MSIT; RS-2026-25482692) and the Global-LAMP Program funded by the Ministry of Education (RS-2023-00301976).
This research used data obtained with the Dark Energy Spectroscopic Instrument (DESI). DESI construction and operations is managed by the Lawrence Berkeley National Laboratory. This material is based upon work supported by the U.S. Department of Energy, Office of Science, Office of High-Energy Physics, under Contract No. DE–AC02–05CH11231, and by the National Energy Research Scientific Computing Center, a DOE Office of Science User Facility under the same contract. Additional support for DESI was provided by the U.S. National Science Foundation (NSF), Division of Astronomical Sciences under Contract No. AST-0950945 to the NSF’s National Optical-Infrared Astronomy Research Laboratory; the Science and Technology Facilities Council of the United Kingdom; the Gordon and Betty Moore Foundation; the Heising-Simons Foundation; the French Alternative Energies and Atomic Energy Commission (CEA); the National Council of Humanities, Science and Technology of Mexico (CONAHCYT); the Ministry of Science and Innovation of Spain (MICINN), and by the DESI Member Institutions: www.desi.lbl.gov/collaborating-institutions. The DESI collaboration is honored to be permitted to conduct scientific research on I’oligam Du’ag (Kitt Peak), a mountain with particular significance to the Tohono O’odham Nation. Any opinions, findings, and conclusions or recommendations expressed in this material are those of the author(s) and do not necessarily reflect the views of the U.S. National Science Foundation, the U.S. Department of Energy, or any of the listed funding agencies.
Funding for the Sloan Digital Sky Survey V has been provided by the Alfred P. Sloan Foundation, the Heising-Simons Foundation, the National Science Foundation, and the Participating Institutions. SDSS acknowledges support and resources from the Center for High-Performance Computing at the University of Utah. SDSS telescopes are located at Apache Point Observatory, funded by the Astrophysical Research Consortium and operated by New Mexico State University, and at Las Campanas Observatory, operated by the Carnegie Institution for Science. The SDSS web site is \url{www.sdss.org}.
SDSS is managed by the Astrophysical Research Consortium for the Participating Institutions of the SDSS Collaboration, including the Carnegie Institution for Science, Chilean National Time Allocation Committee (CNTAC) ratified researchers, Caltech, the Gotham Participation Group, Harvard University, Heidelberg University, The Flatiron Institute, The Johns Hopkins University, L'Ecole polytechnique f\'{e}d\'{e}rale de Lausanne (EPFL), Leibniz-Institut f\"{u}r Astrophysik Potsdam (AIP), Max-Planck-Institut f\"{u}r Astronomie (MPIA Heidelberg), Max-Planck-Institut f\"{u}r Extraterrestrische Physik (MPE), Nanjing University, National Astronomical Observatories of China (NAOC), New Mexico State University, The Ohio State University, Pennsylvania State University, Smithsonian Astrophysical Observatory, Space Telescope Science Institute (STScI), the Stellar Astrophysics Participation Group, Universidad Nacional Aut\'{o}noma de M\'{e}xico, University of Arizona, University of Colorado Boulder, University of Illinois at Urbana-Champaign, University of Toronto, University of Utah, University of Virginia, Yale University, and Yunnan University.  
This publication makes use of data products from the Two Micron All Sky Survey, which is a joint project of the University of Massachusetts and the Infrared Processing and Analysis Center/California Institute of Technology, funded by the National Aeronautics and Space Administration and the National Science Foundation.
The data processing and analysis for this work were performed using the Python programming language, specifically relying on the NumPy \citep{NumPy2020}, Matplotlib \citep{Matplotlib2007}, SciPy \citep{SciPy2020}, Pandas \citep{Pandas2010}, and PyTorch \citep{PyTorch2019} libraries.

%%% APPENDICES (IF ANY) %%%%%%%%%%%%%%%%%%%%%%%%%%%%%%%%%%%%%%%%%%%%%

\begin{table*}
    \centering
    \caption{The Enhanced Catalog of \textit{Gaia} DR3 \texttt{galaxy\_candidates} table incorporating redshifts.}
    \begin{tabular}{ccccccc}
        \toprule
        \texttt{SOURCE\_ID} & \texttt{ra} & \texttt{dec} & \texttt{z} & \texttt{z\_err} & \texttt{z\_flag} & \texttt{P0} \\ 
        (1)&(2)&(3)&(4)&(5)&(6)&(7) \\ \hline
        99956774643712 & 44.78180 & 0.76345 & 0.1344 & 0.0001 & 0 & 0.0E+00 \\
        107717780103296 & 44.99787 & 0.74426 & 0.2288 & 0.0562 & 1 & 5.0E-04 \\
        115006339387776 & 45.21483 & 0.80737 & 0.0435 & 0.0001 & 0 & 0.0E+00 \\
        132495447092608 & 45.04268 & 0.94843 & 0.1074 & 0.0001 & 0 & 0.0E+00 \\
        178331337892736 & 46.15309 & 1.00262 & 0.1549 & 0.0001 & 0 & 0.0E+00 \\
        219769182166656 & 45.35110 & 1.17301 & 0.0715 & 0.0001 & 0 & 0.0E+00 \\
        228805793364352 & 45.18033 & 1.06907 & 0.0712 & 0.0001 & 0 & 0.0E+00 \\
        255915626944256 & 45.90495 & 1.40624 & 0.1283 & 0.0453 & 1 & 3.0E-04 \\
        260279313915008 & 45.85534 & 1.49816 & 0.1425 & 0.0704 & 1 & 5.8E-04 \\
        282441345261824 & 44.34108 & 0.68662 & 0.1353 & 0.0001 & 0 & 0.0E+00 \\
        301197469415552 & 44.11304 & 0.87573 & 0.0229 & 0.0001 & 0 & 0.0E+00 \\
        397443389394944 & 44.42495 & 1.45023 & 0.1767 & 0.0001 & 0 & 0.0E+00 \\
        401016802379392 & 44.45173 & 1.50956 & 0.0932 & 0.0001 & 0 & 0.0E+00 \\
        423281912785920 & 45.04341 & 1.30466 & 0.1795 & 0.0485 & 1 & 5.0E-04 \\
        450181292249472 & 44.94278 & 1.58819 & 0.2212 & 0.0001 & 0 & 0.0E+00 \\
        462245856087168 & 45.27223 & 1.61634 & 0.1299 & 0.0001 & 0 & 0.0E+00 \\
        472171524900096 & 45.38688 & 1.81252 & 0.2489 & 0.0001 & 0 & 0.0E+00 \\
        482655540493952 & 45.22662 & 1.89450 & 0.0764 & 0.0001 & 0 & 0.0E+00 \\
        568795404660736 & 46.23827 & 1.33917 & 0.1788 & 0.0644 & 1 & 3.3E-04 \\
        578106893726848 & 46.48303 & 1.53391 & 0.1571 & 0.0418 & 1 & 5.8E-04 \\
        &&&$\vdots$&&& \\ 
        \bottomrule
    \end{tabular}
    \begin{tablenotes}
        \item \textbf{Note.} The first 20 rows of the \textit{Gaia} redshift catalog constructed in this work. (1) \textit{Gaia} DR3 source id. (2) Right ascension in degrees. (3) Declination in degrees. (4) Redshift. (5) Uncertainty in redshift (6) Redshift flag. 0: spectroscopic, 1: photometric, 2: insufficient info for $P_0$. (7) $P_0$, \textbf{the the risk of trusting it as a genuine extragalactic source or trusting its photometric redshift.} $P_0=-9$ for sources with \texttt{z\_flag}=2
        \item (The full table is available in its entirety in machine-readable form.)
    \end{tablenotes} 
    \label{GaiaRedshiftCatalog}
\end{table*}

\bibliographystyle{jkas}   % or the style required by JKAS
\bibliography{references}

\begin{thebibliography}{}
\expandafter\ifx\csname natexlab\endcsname\relax\def\natexlab#1{#1}\fi
\providecommand{\url}[1]{\href{#1}{#1}}
\providecommand{\dodoi}[1]{doi:~\href{http://doi.org/#1}{\nolinkurl{#1}}}
\providecommand{\doeprint}[1]{\href{http://ascl.net/#1}{\nolinkurl{http://ascl.net/#1}}}
\providecommand{\doarXiv}[1]{\href{https://arxiv.org/abs/#1}{\nolinkurl{https://arxiv.org/abs/#1}}}
\providecommand{\dodoilink}[2]{\href{http://doi.org/#1}{#2}}
\providecommand{\doadslink}[2]{\href{#1}{#2}}

\bibitem[{Abdurro'uf {et~al.}(2022)Abdurro'uf, Accetta, Aerts, \& {et
  al.}}]{SDSSDR17}
Abdurro'uf, A., Accetta, K., Aerts, C., \& {et al.} 2022, ApJS, 259, 35

\bibitem[{Bahk \& Hwang(2024)}]{Bahk2024}
Bahk, H., \& Hwang, H.~S. 2024, ApJS, 272, 7

\bibitem[{Ball {et~al.}(2007)Ball, Brunner, Myers, \& {et al.}}]{Ball2007}
Ball, N.~M., Brunner, R.~J., Myers, A.~D., \& {et al.} 2007, ApJ, 663, 774

\bibitem[{Brammer {et~al.}(2008)Brammer, van Dokkum, \& Coppi}]{Brammer2008}
Brammer, G.~B., van Dokkum, P.~G., \& Coppi, P. 2008, ApJ, 686, 1503

\bibitem[{Carliles {et~al.}(2010)Carliles, Budav{\'a}ri, Heinis, \& {et
  al.}}]{Carliles2010}
Carliles, S., Budav{\'a}ri, T., Heinis, S., \& {et al.} 2010, ApJ, 712, 511

\bibitem[{Cavuoti {et~al.}(2017)Cavuoti, Tortora, Brescia, Longo, \& {et
  al.}}]{Cavuoti2017}
Cavuoti, S., Tortora, C., Brescia, M., Longo, G., \& {et al.} 2017, MNRAS, 466,
  2039

\bibitem[{Colless {et~al.}(2001)Colless, Dalton, Maddox, \& {et
  al.}}]{Colless2001}
Colless, M., Dalton, G., Maddox, S., \& {et al.} 2001, MNRAS, 328, 1039

\bibitem[{Collister \& Lahav(2004)}]{Collister2004}
Collister, A.~A., \& Lahav, O. 2004, PASP, 116, 345

\bibitem[{DESI Collaboration}(2026)]{DESI2025}
{DESI Collaboration (Abdul-Karim, M., Adame, A. G., et al.)}. 2026, AJ, 171,
  285

\bibitem[{Dey {et~al.}(2019)Dey, Schlegel, Lang, \& {et al.}}]{Dey2019}
Dey, A., Schlegel, D.~J., Lang, D., \& {et al.} 2019, AJ, 157, 168

\bibitem[{Falco {et~al.}(1999)Falco, Kurtz, Geller, \& {et al.}}]{Falco1999}
Falco, E.~E., Kurtz, M.~J., Geller, M.~J., \& {et al.} 1999, PASP, 111, 438

\bibitem[{Firth {et~al.}(2003)Firth, Lahav, \& Somerville}]{Firth2003}
Firth, A., Lahav, O., \& Somerville, R. 2003, MNRAS, 339, 1195

\bibitem[{Gaia Collaboration}(2023a)]{GaiaCollaboration2023}
{Gaia Collaboration (Bailer-Jones, C. A. L., Teyssier, D., et al.)}. 2023a,
  A\&A, 674, A41

\bibitem[{Gaia Collaboration}(2023b)]{Carnerero2023}
{Gaia Collaboration (Carnerero, M. I., Raiteri, C. M., Rimoldini, L., et al.)}.
  2023b, A\&A, 674, A24

\bibitem[{Gaia Collaboration}(2016)]{GaiaCollaboration2016}
{Gaia Collaboration (Prusti, T., de Bruijne, J. H. J., et al.)}. 2016, A\&A,
  595, A1

\bibitem[{Gaia Collaboration}(2022)]{GaiaAGN2022}
{Gaia Collaboration (Rimoldini, L., Audard, M., Holl, B., et al.)}. 2022, A\&A,
  674, A1

\bibitem[{Glorot \& Bengio(2010)}]{GlorotBengio2010}
Glorot, X., \& Bengio, Y. 2010, in JMLR Workshop and Conference Proceedings,
  Vol.~9, Proceedings of the 13th International Conference on Artificial
  Intelligence and Statistics, 249

\bibitem[{Green(2018)}]{Green2018}
Green, G.~M. 2018, JOSS, 3, 695

\bibitem[{Harris {et~al.}(2020)Harris, Millman, van~der Walt, \& {et
  al.}}]{NumPy2020}
Harris, C.~R., Millman, K.~J., van~der Walt, S.~J., \& {et al.} 2020, Nature,
  585, 357

\bibitem[{Heintz {et~al.}(2020)Heintz, Fynbo, Geier, \& {et al.}}]{Heintz2020}
Heintz, K.~E., Fynbo, J. P.~U., Geier, S.~J., \& {et al.} 2020, A\&A, 644, A17

\bibitem[{Huchra {et~al.}(2012)Huchra, Macri, Masters, \& {et al.}}]{2MRS2012}
Huchra, J.~P., Macri, L.~M., Masters, K.~L., \& {et al.} 2012, ApJS, 199, 26

\bibitem[{Hunter(2007)}]{Matplotlib2007}
Hunter, J.~D. 2007, CSE, 9, 90

\bibitem[{Hwang {et~al.}(2010)Hwang, Elbaz, Lee, \& {et al.}}]{Hwang2010}
Hwang, H.~S., Elbaz, D., Lee, J.~C., \& {et al.} 2010, A\&A, 522, A33

\bibitem[{Hwang {et~al.}(2014)Hwang, Geller, Diaferio, Rines, \&
  Zahid}]{Hwang2014}
Hwang, H.~S., Geller, M.~J., Diaferio, A., Rines, K.~J., \& Zahid, H.~J. 2014,
  ApJ, 797, 106

\bibitem[{Hwang {et~al.}(2016)Hwang, Geller, Park, \& {et al.}}]{Hwang2016}
Hwang, H.~S., Geller, M.~J., Park, C., \& {et al.} 2016, ApJ, 818, 173

\bibitem[{Jarrett {et~al.}(2000)Jarrett, Chester, Cutri, Schneider, Skrutskie,
  \& Huchra}]{2MASSXSC}
Jarrett, T.~H., Chester, T., Cutri, R.~M., {et~al.} 2000, AJ, 119, 2498

\bibitem[{Jones {et~al.}(2004)Jones, Saunders, Colless, \& {et
  al.}}]{Jones2004}
Jones, D.~H., Saunders, W., Colless, M., \& {et al.} 2004, MNRAS, 355, 747

\bibitem[{Jordi {et~al.}(2010)Jordi, Gebran, Carrasco, \& {et al.}}]{Jordi2010}
Jordi, C., Gebran, M., Carrasco, J.~M., \& {et al.} 2010, A\&A, 523, A48

\bibitem[{Kim {et~al.}(2025)Kim, Sohn, Hwang, \& {et al.}}]{Kim2025}
Kim, T., Sohn, J., Hwang, H.~S., \& {et al.} 2025, ApJS, 277, 41

\bibitem[{Kingma \& Ba(2015)}]{Adam}
Kingma, D.~P., \& Ba, J. 2015, in ICLR

\bibitem[{Kwon {et~al.}(2026)Kwon, Hwang, Lee, \& {et al.}}]{Kwon2026}
Kwon, M., Hwang, H.~S., Lee, J.~C., \& {et al.} 2026, JKAS, 59, 1

\bibitem[{Lee \& Shin(2021)}]{MBRNN}
Lee, J., \& Shin, M. 2021, AJ, 162, 297

\bibitem[{McKinney(2010)}]{Pandas2010}
McKinney, W. 2010, in Proceedings of the 9th Python in Science Conference, ed.
  S.~van~der Walt \& J.~Millman, 51

\bibitem[{Paszke {et~al.}(2019)Paszke, Gross, Massa, \& {et al.}}]{PyTorch2019}
Paszke, A., Gross, S., Massa, F., \& {et al.} 2019, in Advances in Neural
  Information Processing Systems 32, 8024

\bibitem[{Peacock {et~al.}(2001)Peacock, Cole, Norberg, Baugh, Bland-Hawthorn,
  Bridges, Cannon, \& {et al.}}]{Peacock2001}
Peacock, J.~A., Cole, S., Norberg, P., {et~al.} 2001, Nature, 410, 169

\bibitem[{{Planck Collaboration (Greiss, S., Steeghs, D., et
  al.)}(2014)}]{Greiss2014}
{Planck Collaboration (Greiss, S., Steeghs, D., et al.)}. 2014, A\&A, 571, A16

\bibitem[{Quintana \& Ramirez(1995)}]{Quintana1995}
Quintana, H., \& Ramirez, A. 1995, ApJS, 96, 343

\bibitem[{Riello {et~al.}(2021)Riello, De~Angeli, Evans, \& {et
  al.}}]{GaiaEDR3Phot}
Riello, M., De~Angeli, F., Evans, A., \& {et al.} 2021, A\&A, 649, A3

\bibitem[{Roche {et~al.}(2024)Roche, Necib, Lin, Ou, \& Nguyen}]{Roche2024}
Roche, C., Necib, L., Lin, T., Ou, X., \& Nguyen, T. 2024, ApJ, 972, 70

\bibitem[{Salvato {et~al.}(2019)Salvato, Ilbert, \& Hoyle}]{Salvato2019}
Salvato, M., Ilbert, O., \& Hoyle, B. 2019, Nature Astronomy, 3, 212

\bibitem[{Saunders {et~al.}(2000)Saunders, Sutherland, Maddox, \& {et
  al.}}]{Saunders2000}
Saunders, W., Sutherland, W.~J., Maddox, S.~J., \& {et al.} 2000, MNRAS, 317,
  55

\bibitem[{Schlafly {et~al.}(2019)Schlafly, Meisner, \& Green}]{unWISE2019}
Schlafly, E.~F., Meisner, A.~M., \& Green, G.~M. 2019, ApJS, 240, 30

\bibitem[{Skrutskie {et~al.}(2006)Skrutskie, Cutri, Stiening, \& {et
  al.}}]{2MASS2006}
Skrutskie, M.~F., Cutri, R.~M., Stiening, R., \& {et al.} 2006, AJ, 131, 1163

\bibitem[{Virtanen {et~al.}(2020)Virtanen, Gommers, Oliphant, \& {et
  al.}}]{SciPy2020}
Virtanen, P., Gommers, R., Oliphant, T.~E., \& {et al.} 2020, NatMe, 17, 261

\bibitem[{Wright {et~al.}(2010)Wright, Eisenhardt, Mainzer, \& {et
  al.}}]{Wright2010}
Wright, E.~L., Eisenhardt, P. R.~M., Mainzer, A.~K., \& {et al.} 2010, AJ, 140,
  1868

\end{thebibliography}

%%% CALL LIST OF REFERENCES (natbib STYLE) %%%%%%%%%%%%%%%%%%%%%%%%%%
% \bibliography{jkas-sample}

%\begin{thebibliography}{}

\end{document}